\documentclass{article}
\usepackage[T2A]{fontenc}
\usepackage[cp1251]{inputenc}
\usepackage[english]{babel}

\usepackage[tbtags]{amsmath}
\usepackage{amssymb,amsmath}

\newtheorem{theorem}{Theorem}

\begin{document}

\title{\bf\large\MakeUppercase{On critical points of the objective functional for maximization of qubit observables}}

\makeatletter
\renewcommand{\@makefnmark}{}
\makeatother
\footnotetext{This work is supported by the Russian Science Foundation under grant 14-50-00005.}
\author{Alexander N. Pechen\footnote{\texttt{pechen@mi.ras.ru}}~~and Nikolay B. Il'in\footnote{\texttt{ilyn@mi.ras.ru}}\\
\textit{Steklov Mathematical Institute of Russian Academy of Sciences,}\\ \textit{Gubkina 8, Moscow 119991, Russia}
}
\date{}

\maketitle

An isolated from the environment $n$-level controlled quantum system is described by the Schr\"odinger equation for unitary evolution operator $U_t$:
\begin{equation}\label{e00}
i\frac{dU_t}{dt}=(H_0+f(t)V)U_t,\qquad U_{t=0}=\mathbb I\,.
\end{equation}
Here $H_0$ and  $V$  are $n\times n$  Hermitian matrices and $f(t)\in L^1([0,T];\mathbb R)$ is the control. We assume that $[H_0,V]\neq 0$. The goal is to find a control $f(t)$ which maximizes the objective functional ${\cal J}_A[f] = {\rm Tr}(U_T\rho_0U^{\dagger}_T A)$, where  $\rho_0$ is the initial density matrix of the system, $A$ is a Hermitian matrix and $T>0$ is the final time. The functional ${\cal J}_A$  describes the average value of an observable  $A$ at time $T$. An important problem in quantum control is the analysis of local maxima and minima (called traps) of the objective functional~\cite{Rabitz,RabitzHo,PechenTannor} because traps, if they exist, can be obstacles to find globally optimal solutions. The absence of traps in typical  control problems for quantum systems was suggested in~\cite{Rabitz,RabitzHo}. The absence of traps for two-level quantum systems for sufficiently large $T$ was proved in~\cite{Pechen, Mian}. The influence of constraints in the controls on the appearance of traps for two-level Landau-Zener system was investigated in~\cite{Zhdanov}. For constrained control of this system with sufficiently large $T$ only a special kind of traps may exist. These traps can be eliminated by a modification of gradient search algorithm so that in this sense the quantum control landscape practically appears as trap-free~\cite{Zhdanov}. In~\cite{Mian} the following statement is proved.
\begin{theorem}\label{th0}
If ${\rm Tr} V=0$ and $T\geq T_0$, where $T_0=\pi/\|H_0-(1/2){\rm Tr}
H_0+f_0V\|$ and $f_0=-{\rm Tr}(H_0V)/{\rm Tr}(V^2)$, then all maxima of the objective functional  ${\cal J}_A$ are global.
\end{theorem}
From the analysis of work~\cite{Mian} it follows that any $f\neq f_0$ is not a trap for any $T>0$. Hence, only the control $f=f_0$ may be a trap for small $T$.

Below we consider the special case of the Schr\"odinger equation~(\ref{e00}) of the form:
\begin{equation}\label{e01}
i\frac{dU_t}{dt}=(\sigma_z+f(t)(v_x\sigma_x+v_y\sigma_y))U_t\,.
\end{equation}
Here $\sigma_x,\sigma_y,\sigma_z$ are the Pauli matrices. In this case $f_0=0$ and $T_0=\pi$. In~\cite{Mian2} it was proved that traps for this system may appear for sufficiently small  $T$, but only if the vectors ${\bf r}^0={\rm Tr}(\rho_0\boldsymbol{\sigma})$,  ${\bf a}={\rm Tr}(A\boldsymbol{\sigma})$ and ${\bf v}=1/2{\rm Tr}(V\boldsymbol{\sigma})$ ($\boldsymbol{\sigma}=(\sigma_x,\sigma_y,\sigma_z)$) belong to the plane which is orthogonal to the vector ${\bf h}_0={\rm Tr}(H_0\boldsymbol{\sigma})$. In ~\cite{Mian2} the following statement was proved.
\begin{theorem}\label{th1}
If $[({\bf v}\times{\bf r}^0)_z\cos 2T + ({\bf v}\cdot{\bf r}^0)\sin 2T]({\bf v}\times{\bf a})_z>0$,  $({\bf r}^0\times {\bf a})_z\cos 2T<({\bf r}^0\cdot{\bf a})\sin 2T$ and the vectors ${\bf r}^0$,  ${\bf a}$ and ${\bf v}$ belong to the plane which is orthogonal to the vector ${\bf h}_0$, then there exists such $T_*$ that for all $T\leq T_*$ the control $f(t)=0$ is a traps for maximization of the objective functional ${\cal J}_A$.
\end{theorem}
Despite this result, the problem of  analysis of traps for two-level systems still is not completely solved. So, for the system~(\ref{e01}) according to theorem~\ref{th0} there are no traps for all $T\geq T_0=\pi$ and according to theorem~\ref{th1}  traps exist for all $T\leq T_*$, but the value of $T_*$ is not specified. Below we prove that  the lower bound on $T$ for the absence of traps can be reduced.
\begin{theorem}\label{th2}
For $T>\pi/2$ the control $f=0$ is not a trap for maximization of the objective functional ${\cal J}_A$ for the system~(\ref{e01}).
\end{theorem}
In order to prove that the control $f=0$ is not a trap, we will prove that if $f=0$ is a critical point then it is either a global extremum or a saddle point. In the latter case it is sufficient to verify that the Hessian at $f=0$ has eigenvalues of opposite signs. In ~\cite{Mian2} it is proved that the control $f=0$ can be a trap only if the vectors ${\bf r}^0$, ${\bf a}$ and ${\bf v}$ belong to the plane which is orthogonal to the vector ${\bf h}_0$. Under this condition on the vectors ${\bf r}^0$, ${\bf a}$ and ${\bf v}$
in~\cite{Mian} the following expression was obtained for Hessian at the point $f=0$ in terms of vectors  ${\bf r}={\rm Tr}(U_T \rho U_T^\dagger \boldsymbol{\sigma})$ and ${\bf r}_{t}=\sin(2t-\phi){\bf e}_x+\cos(2t-\phi){\bf e}_y$, where $\phi=\arctan(v_y/v_x)$:
\begin{eqnarray}\label{Hess}
(f,Hf)&=&\int\limits_0^T\int\limits_0^T{\rm Hess}_{f}{\cal J}_A(t_2,t_1)f(t_1)f(t_2)dt_1dt_2,\nonumber\\
{\rm Hess}_{f}{\cal J}_A(t_2,t_1)&=&\left\{\begin{matrix}
-\dfrac{v^{2}}{4}({\bf r}\cdot{\bf r}_{t_2})({\bf a}\cdot{\bf r}_{t_1}), \, t_2\geqslant t_1 \vspace{1mm} \\
-\dfrac{v^{2}}{4}({\bf r}\cdot{\bf r}_{t_1})({\bf a}\cdot{\bf r}_{t_2}), \, t_2< t_1
\end{matrix}\right.
\end{eqnarray}
Define the function
\begin{equation}
\delta_{\varepsilon} (t)=\begin{cases} 0, & |t|\geqslant \dfrac{\varepsilon}{2}\\
\dfrac{1}{\varepsilon}, & |t|\leqslant \dfrac{\varepsilon}{2}
\end{cases}
\end{equation}
For all $f\in {\rm C}[0,T]$ and for all $t\in(\varepsilon / 2,~T-\varepsilon / 2)$, $\varepsilon<T$ we have $\int\limits_0^T \delta_{\varepsilon} (\tau-t) f(\tau)d\tau=f(t)+O(\varepsilon)$.
Let $f_{\lambda}(t)=\delta_{\varepsilon} (t-\lambda)$, $\varepsilon / 2<\lambda<T-\varepsilon / 2$.  Substitution of the function $f_{\lambda}(t)$ in the expression for Hessian ($\ref{Hess}$) gives $(f_{\lambda},Hf_{\lambda})=-{v^{2}}({\bf r}\cdot{\bf r}_{\lambda})({\bf a}\cdot{\bf r}_{\lambda})/{4}+O(\varepsilon)$, where
\begin{equation}\label{e3}
({\bf r}\cdot{\bf r}_{\lambda})({\bf a}\cdot{\bf r}_{\lambda})=\dfrac{|{\bf r}||{\bf a}|}{2}\bigg(\cos(\phi_1-\phi_2)-\cos(4\lambda-2\phi+\phi_1+\phi_2)\bigg)
\end{equation}
Here $\phi_1=\arctan(r_y/r_x)$ and $\phi_2=\arctan(a_y/a_x)$. If $\cos(\phi_1-\phi_2)\neq\pm 1$ then it follows from~(\ref{e3})
that the function $({\bf r}\cdot{\bf r}_{\lambda})({\bf a}\cdot{\bf r}_{\lambda})$ has values of opposite signs on the interval $[0,\pi/2]$.
In this case we choose $\lambda_1$ so that $({\bf r}\cdot{\bf r}_{\lambda_1})({\bf a}\cdot{\bf r}_{\lambda_1})<0$ and $\lambda_2$ so that $({\bf r}\cdot{\bf r}_{\lambda_2})({\bf a}\cdot{\bf r}_{\lambda_2})>0$. Then we choose $\varepsilon$ small enough so that the sign of $(f_{\lambda},Hf_{\lambda})$ is the same as the sign of  $({\bf r}\cdot{\bf r}_{\lambda})({\bf a}\cdot{\bf r}_{\lambda})$. Then  $(f_{\lambda_1},Hf_{\lambda_1})>0$ and $(f_{\lambda_2},Hf_{\lambda_2})<0$, i.e. Hessian at the control $f=0$ has eigenvalues of opposite signs. This means that the control $f=0$ is a saddle point. If $\cos(\phi_1-\phi_2)=\pm 1$ then the vectors ${\bf r}$ and ${\bf a}$ are parallel. Hence, we have ${\cal J}_A[0]=(1\pm|{\bf a}||{\bf r}|)/2$ that is a global extremum of the objective functional.

\end{document}